\documentstyle[12pt,epsfig]{article}
\def\b{\bar}
\def\d{\partial}
\def\D{\Delta}
\def\cD{{\cal D}}
\def\cK{{\cal K}}

\def\G{\Gamma}
\def\l{\lambda}

\def\m{\mu}
\def\n{\nu}

\def\q{\b q}

\def\t{\tau}

\def\~{\tilde}
\def\h{\eta}

\def\bY3{\bar Y_{,3}}
\def\Y3{Y_{,3}}
\def\z{\zeta}
\def\Z{{\b\zeta}}
\def\Y{{\bar Y}}
\def\cZ{{\bar Z}}
\def\`{\dot}
\def\be{\begin{equation}}
\def\ee{\end{equation}}
\def\bea{\begin{eqnarray}}
\def\eea{\end{eqnarray}}

\def\fn{\footnote}

\def\cF{{\cal F}}

\def\mn{{\mu\nu}}

\begin{document}
\title{`Alice' String as Source of the Kerr Spinning Particle}

\author{Alexander Burinskii\\
Gravity Research Group, NSI Russian
Academy of Sciences\\
B. Tulskaya 52, 115191 Moscow, Russia}
\maketitle

\begin{abstract}
\noindent Kerr geometry has twofoldedness which can be cured by a
truncation of the `negative' sheet of metric. It leads to the
models of disk-like sources of the Kerr solution and to a class of
disk-like or bag-like models of the Kerr spinning particle. There
is an alternative way: to retain the `negative' sheet as the sheet
of advanced fields. In this case the source of spinning particle
is the Kerr singular ring which can be considered as a twofold
`Alice' string. This string can have electromagnetic excitations
in the form of traveling waves generating spin and mass of the
particle. Model of this sort was suggested in 1974 as a `microgeon
with spin'. Recent progress in the obtaining of the nonstationary
and radiating Kerr solutions enforces us to return to this model
and to consider it as a model for the light spinning particles. We
discuss here the real and complex Kerr geometry and some unusual
properties of the oscillating solutions in the model of `Alice'
string source.
\end{abstract}

\section{Introduction}

It is known, that Kerr geometry displays some properties of spinning
particle and diverse classical models of spinning particle were considered
on the base of Kerr geometry \cite{Car,Isr,Bur0,Lop,IvBur}.
Since the Kerr geometry has a topological twofoldedness of space-time,
there appears an alternative: either to remove this twofoldedness or to give it
a physical interpretation. The both approaches have paid attention and were
considered in literature, it seems
however that the final preference for the solution of this problem is not
formed yet. It is also possible that the both versions of the Kerr source
can be valid and applicable for different models.
The approach which was the most popular last time is the truncation of the
negative sheet of the Kerr geometry.
It leads to the appearance of a (relativistically rotating) disk-like source of
 the Kerr solution \cite{Isr} and to a class of the disk-like \cite{Lop} or
 bag-like \cite{Bag,BEHM} models of the Kerr spinning particle.

 Alternative way is to retain the negative sheet treating it
as the sheet of advanced fields. In this case the source of spinning particle
turns out to be the Kerr singular ring and its electromagnetic excitations
in the form of traveling waves generating spin and mass of the particle.
Model of this sort can be considered as a twofold `Alice' string \cite{Wit}
and was suggested in 1974 as a model of `microgeon with spin'\cite{Bur0}.
Recent progress in the obtaining of the nonstationary and radiating Kerr
solutions \cite{Bur-nst} enforced us to return to this model and to consider
it as a plausible classical model for the light spinning particles.

In this paper we discuss the real and complex Kerr structures of the Kerr
geometry and a way for obtaining the exact solutions for microgeon.
We display also the CPT-invariance of Kerr geometry and
consider the physical interpretation of the negative sheet leading to
some unusual properties of the oscillating solutions of this kind.

In this paper we use the Kerr-Schild approach to
 the Kerr geometry which is based on
the Kerr-Schild form of metrics $g^\mn=\eta ^\mn - 2h k^\m k^\n $,
where $\eta^\mn$ is the auxiliary Minkowski background. It allows to
give an exact meaning to the complex representation of the Kerr geometry and
simplifies the treatment of the complex null cones that is necessary to
consider the complex retarded-time construction for oscillating solutions.

\section{Main peculiarities of the real  Kerr
geometry}

{\bf The Kerr singular ring} is one of the most remarkable peculiarities of
the Kerr solution.  It is a branch line of space on two sheets: "negative"
and
"positive" where the fields change their signs and directions.

{\bf The Kerr twisting PNC} is the second remarkable structure of the Kerr
geometry.  It is described by a vector field $k^\m$ which
determines the Kerr-Schild ansatz for metric
\be g_{\m\n} = \h_{\m\n} + 2 h
k_{\m} k_{\n}, \label{ksa} \ee where
$ \h_\mn $ is metric of auxiliary
Minkowski space-time and
\be h= \frac {mr-e^2/2} {r^2 + a^2 \cos^2 \theta}.
\ee
This is a remarkable simple form showing that all the complicatedness
of the Kerr solution is included in the form of the field $k_\m (x)$
which is tangent to the Kerr PNC.
This form shows also that metric is singular at $r=\cos\theta=0$, that are
the focal points of the oblate spheroidal coordinate system.

\begin{figure}[ht]
\centerline{\epsfig{figure=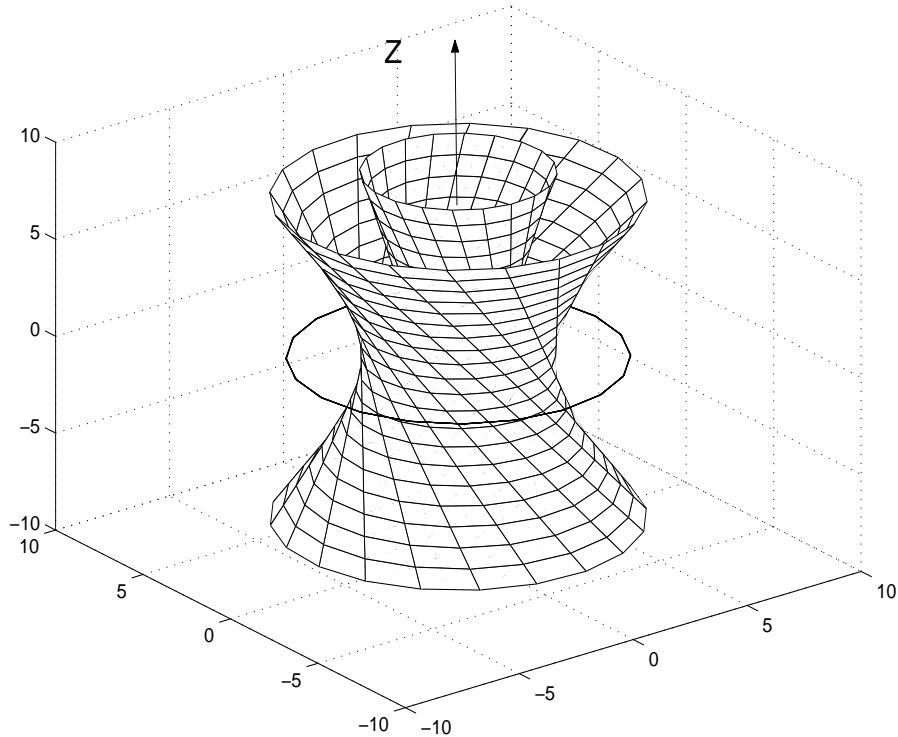,height=10cm,width=12cm}}
\caption{The Kerr singular ring and 3-D section of the Kerr principal null
congruence (PNC). Singular ring is a branch line of space, and PNC
propagates from ``negative'' sheet of the Kerr space to ``positive '' one,
covering the space-time twice. } \end{figure}

Field $k^\m$ is null
with respect to $\h_\mn$ as well as with respect
to the metric $g_\mn$.
The Kerr singular ring and a part of the Kerr PNC are shown on the
fig.1.
The Kerr PNC consists of the linear generators of the surfaces $\theta
 =const$. The shown on the fig.1 region $z<0$ corresponds to a ``negative''
sheet of space ($r<0$)  where we set the null rays to be ``in''-going.

Twisting vortex of the null rays propagates through the singular ring
$r=\cos\theta=0$ and get ``out'' on the ``positive'' sheet of space ($z>0$).
Indeed, the Kerr congruence covers the space-time twice, and this picture
shows only the half of PNC corresponding to $0>\theta>\pi /2$.  It has to
be completed by the part for $\pi/2 >\theta>\pi$ which is described by
another system of the linear generators (having opposite twist). Two PNC
directions for the each point $x^\m\in M^4$
correspond to the known twofoldedness of the Kerr geometry and
to the algebraically degenerated metrics of type D.

As it is explicitly seen from the expression for $h$, the Kerr gravitational
field has twovaluedness, $h(r) \ne h(-r)$, and so do also the other fields on
the Kerr background. The oblate coordinate system turns out to be very useful
since it covers also the space twice, for $r>0$ and $r<0$ with the
branch line on the Kerr singular ring.

\section{Kerr singular ring as the `Alice' string}
In the case $e^2 +a^2 >>m^2$, corresponding to parameters of
elementary particles, the horizons of the Kerr-Newman solution
disappear and the Kerr singular ring turns out to be naked.
The naked Kerr singular ring was considered in the model
of spinning particle - microgeon  \cite{Bur0} -
as a waveguide providing a circular propagation of an electromagnetic or
fermionic wave excitation. Twofoldedness of the Kerr geometry admits
integer and half integer wave excitations with
 $n=2\pi a/\lambda$ wave periods  on the Kerr ring of radius
 $a$. It is consistent with the corresponding values of the
angular momentum $J=n \hbar $ and mass $m$ in accordance with the
main relation for the Kerr parameters $m= J/a$. Radius of the Kerr
ring $a=n\hbar/m$ turns out to be of the order of the
corresponding Compton size. It was recognized soon \cite{IvBur}
that this construction can be considered as a closed string with a
spin excitations. This proposal was natural since singular lines
were used in many physical models of the dual relativistic
strings. The most well known example of this kind is the
Nielsen-Olesen model representing a vortex line in superconductor,
another example is the Witten superconducting cosmic string
\cite{Wit}.

The Kerr ring displays a few stringy properties. First, the
Compton size of Kerr ring $a$ is a typical size of the
relativistic string excitations. Because of that the radius of
interaction of the Kerr spinning particle is not determined by
mass parameter $m$, as it has the place for other gravitational
solutions, but it is extended to the Compton distances $\sim a$.
In the same time the contact stringy character of interaction is
provided with a very small effective cross-section. If we assume
the existence of a stringy tension $T$, so that $E=m=2\pi Ta$,
then, in the combination with the Kerr relation  $J=ma$, one
obtains the Regge relation  $J=(2\pi T)^{-1}m^2$. Finally, as it
was shown in \cite{BurSen} by the analysis of the axidilatonic
generalization of the Kerr solution \cite{Sen}, the field near the
Kerr singular ring of this solution is similar to the field around
a heterotic string.  Note also, that the Kerr ring is a
relativistic light-like object, that can be seen from the analysis
of the Kerr null congruence near the ring. The light-like rays of
the Kerr congruence are tangent to the ring, as it is shown on the
Fig.2. In stringy terms the Kerr string contains only modes of one
(say `left') direction. However, the equations for the usual
bosonic closed strings do not admit such solutions in four
dimensions. There are a few ways to avoid this obstacle. One of
them is to assume that the model contains the right modes too, but
they are moving in the fifth compactified direction. Another way
is to consider the traveling waves as the missing `right'
excitations.

In this connection one should mention one more
stringy structure of the Kerr geometry, complex euclidean string \cite{BurStr},
which is related to its complex representation.
\begin{figure}[ht]
\centerline{\epsfig{figure=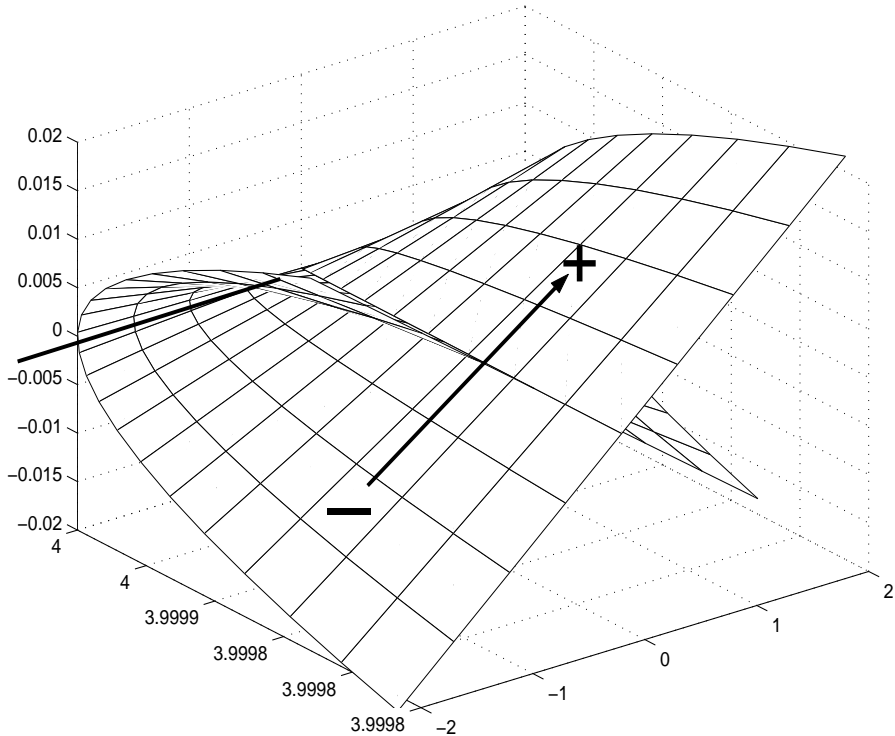,height=10cm,width=12cm}}
\caption{The surface $\Phi=const.$ formed by the light-like generators
of the Kerr principal null congruence (PNC). The Kerr string (fat line) is
tangent to this surface.} \end{figure}

Complex representation of Kerr geometry, has been found useful in
diverse problems \cite{BurStr,BurSup,BurMag}.
In the Kerr-Schild approach it allows one to
get a retarded-time description of the nonstationary Maxwell fields and
twisting algebraically special solutions of the Einstein equations
\cite{Bur-nst}.
Twisting solutions are represented in this approach as the retarded-time
fields which are similar to Lienard-Wiechard fields, however they are
generated by a {\it complex} source moving along a {\it complex world line}
$x_0(\t)$ in complex Minkowski space-time $CM^4$.

 The objects described by the complex world lines occupy an intermediate
position between particle and string.  Like the string they form the
two-dimensional
surfaces or the world-sheets in the space-time \cite{BurStr,OogVaf}.  In many
respects this source is similar to the "mysterious" $N=2$ complex string of
superstring theory \cite{OogVaf}.

It was shown \cite{BurStr}, that analytical complex world lines are the
solutions of the corresponding string equations. Below we shall show
that a given complex world line can control excitations of the
Kerr singular ring. Based on the recent progress in the
obtaining nonstationary Kerr solutions \cite{Bur-nst}, we shall try to get
selfconsistent solutions for traveling waives accompanied by  oscillating
singular ring of the Kerr geometry.

\section{Complex structure of Kerr geometry and related
retarded-time construction}

The light cones emanating from the word line of a source play usually a
central role in the retarded-time constructions where the fields are
defined by the values of a retarded time. In the case of complex
world line, the corresponding light cone has to be complex that
complicates the retarded-time scheme.

\subsection{Appel complex solution.}

There exist the
Newton and Coulomb analogues of the Kerr solution possessing the Kerr
singular ring. It allows one understand the origin of this ring
as well as the complex origin of the Kerr source.
The corresponding Coulomb solution was obtained by Appel still
in 1887 (!) by a method of complex shift \cite{App}.  \par A point-like charge
$e$, placed on the complex z-axis $(x_0,y_0,z_0)= (0,0, ia)$ gives a real
Appel potential \begin{equation} \phi_a = Re \ e/{\tilde r} .  \end{equation}
Here $\tilde r$ is in fact the Kerr complex radial coordinate $\tilde r=
PZ^{-1}= r+ i a \cos\theta$, where $r$ and $\theta$ are the oblate spheroidal
coordinates.  It may be expressed in the usual rectangular Cartesian
coordinates $x,y,z,t$  as
\begin{equation} \tilde r = [(x-x_0)^2 + (y-y_0)^2 +
(z-z_0)^2]^{1/2} = [x^2 + y^2 + (z-ia)^2]^{1/2}.  \end{equation}
Singular line
of the solution corresponds to $r=\cos\theta=0$, and it is seen that the Appel
potential $\phi_a$ is singular at the ring $z=0,\quad x^2+y^2=a^2$.  It was
shown, that this ring is a branch line of space-time for two sheets similarly
to the properties of the Kerr singular ring.  Appel potential describes {\bf
exactly} the e.m. field of the Kerr-Newman solution \cite{Bur0}.

As far the Appel source is shifted to a complex point of space
$(x_o, y_o , z_o ) \rightarrow (0,0,ia)$, it
can be considered as a mysterious "particle" propagating
along a {\it complex world-line} $x_0^\mu (\tau)$ in $CM^4$
and parametrized by a complex time $\tau$.
Complex source of the Kerr-Newman solution
 has just the same origin \cite{LinNew,BurStr} and leads to a complex
retarded-time construction for Kerr geometry.

The appearance of the twisting Kerr
congruence may be understood as a track  of the null planes of the
family of complex light cones emanated from the points of the complex
world line  $x_0^\m(\tau)$ \cite{BKP,BurStr} in the retarded-time
construction. It is very instructive to consider the following splitting
of the complex light cones.

\subsection{Splitting of the complex light cone}

The complex light cone $\cK$ with the vertex at some point $x_0$, written
in spinor form
\be {\cK }= \{x: x =
x^{\m}_{o}(\tau) + \psi ^{A}_{L} \sigma ^{\m}_{A \dot { A}} \tilde{\psi
}^{\dot{A}}_{R} \} \label{sclc} \ee
 may be split  into
two families of null planes: "left" $( \psi _{L}$ =const; $\tilde{\psi
}_{R}$ -var.) and "right"$( \tilde{\psi }_{R}$ =const; $\psi _{L}$ -var.).
These are the only two-dimensional planes which are wholly contained in the
complex null cone.
The rays of the principal null congruence of the Kerr geometry are
the tracks of these complex null planes (right or left) on the real slice of
Minkowski space.

The light cone equation in the Kerr-Schild metric coincides with the
corresponding equation in Minkowski space because the null directions
$k^\m $ are null in both metrics, $g_\mn$ and $\eta _\mn $.
\par

In the null Cartesian coordinates
\bea
2^{1\over2}\z &=& x+iy ,\qquad 2^{1\over2} \Z = x-iy , \nonumber\\
2^{1\over2}u &=& z + t ,\qquad 2^{1\over2}v = z - t . \label{ncc}
\eea

the light cone equation has the form $\z\Z +uv =0$.
As usually, in a complex
extension to $CM^4$ the coordinates $u,v$ have to be considered as complex
and coordinates $\z$ and $\Z$ as independent.
On the real section, in $M^4$, coordinates $u$ and $v$ take the real values
and $\z$ and $\Z$  are complex conjugate.

The known splitting of the light cone on the complex null planes has a close
connection to spinors and twistors. By introducing the projective spinor
parameter $Y = \psi ^{1}/ \psi ^{0}$ the equation of complex light cone
with the vertex at point $x_0$,
\be (\z  -\z _0) (\Z -\Z_0) = - (u - u_0) ( v - v_0) , \ee
splits into two linear equations \footnote{It is a generalization of
the Veblen and Ruse construction \cite{Veb,Rus} which has been used for the
geometrical representation of spinors.}
\bea  \z  - \z _0 &=& Y (v - v_0 ), \\
-Y (\Z - \Z _0 ) &=& (u - u_0 ) \ ,
\label{split1}
\eea
describing the "left" complex null planes
(the  null  rays  in  the real space).  Another splitting
\bea  - \tilde {Y} (\z  - \z _0) &=&  (u - u_0 ), \\
(\Z - \Z_0 ) &=& \tilde{Y} (v - v_0 ) \ ,
\label{split2}
\eea
gives the "right" complex null planes.

Thus, the equations of the "left" null planes (\ref{split1}) can be written in
terms of the three parameters
\be
Y,\qquad \lambda _{1} = \z - Y v,\qquad \lambda _{2} = u + Y \Z ,
\ee
as follows
\be \lambda _{1} = \lambda ^0_{1} , \quad  \lambda _{2} = \lambda ^0_{2} \ ,
 \label{lnp} \ee
where
\be
\lambda ^0_{1} = \z _0 - Y v_0  , \qquad \lambda ^0_{2} = u_0 + Y\Z _0
\label{np} \ee
note the values of these parameters at the point $x_0$.
These three parameters are the projective
twistor variables and very important for further consideration since
the Kerr theorem is
formulated in terms of these parameters. The above
splitting of the complex light cone equation shows explicitly their
origin.
Note also that in the terms of the Kerr-Schild
null tetrad
\begin{eqnarray}
e^1 &=& d \zeta - Y dv, \qquad  e^2 = d \bar\zeta -  \bar Y dv, \nonumber \\
e^3 &=&du + \bar Y d \zeta + Y d \bar\zeta - Y \bar Y dv, \nonumber\\
e^4 &=&dv + h e^3,\label{KSt}
\end{eqnarray}
the projective twistor parameters take the form
\bea
\l_1 &=& x^\m e^1_\mu ,
\nonumber\\
\l _2 &=&x^\m (e^3_\m-\Y e^1_\m ),
\label{ltwea}
\eea
and correspondingly
\bea
\l ^0_1 &=&x_0^\m e^1_\mu ,
\nonumber\\
\l ^0 _2 &=&x_0^\m (e^3_\m-\Y e^1_\m ).
\label{lambx}
\eea
The  ``left'' complex null planes of the complex light cone
at some point $x_0$ can be expressed in terms of the tetrad as follows
\be x_L = x_0(\tau) + \alpha e^1 + \beta e^3 \ ,  \label{L}
\ee
and the null plane equations (\ref{lnp}) follow then from (\ref{L}) and the
tetrad scalar products $e^{1\m} e^1 _\m =e^{1\m} e^3 _\m =
e^{3\m} e^3 _\m =0$.  Similar relations valid also for the ``right'' null
planes with the replacement $e^1 \to e^2$.

The "left" null planes of the
complex light cones form a complex Kerr congruence which generates all the
rays of the principal null congruence on the real space.  The ray with polar
direction $\theta,\phi$ is the real track of the "left" plane corresponding to
$Y = \exp{i \phi} \tan (\theta /2) $ and belonging to the cone which is placed
at the point $x_0$ corresponding to $\sigma = a \cos (\theta).$ The parameter
$\sigma = Im \tau$ has a meaning only in the range $- a \leq \sigma \leq a$
where the cones have real slices.  Thus, the complex world line $x_0(t,\sigma
)$ represents a restricted two-dimensional surface or strip, in complex
Minkowski space, and is really a world-sheet.\fn{It may  be considered  as a
complex open string with a Euclidean parametrization $\tau = t+i\sigma , \bar
\t =t-i\sigma $, and with end points $x_0( t,\pm a)$ \cite{BurStr,OogVaf}.}

 The Kerr congruence arises as the real slice of the family of the
"left" null planes ($Y=const.$) of the complex light cones which vertices
lie on the complex world line $x_0(\tau)$.

The Kerr theorem can be linked to this retarded-time construction.

\section{Kerr theorem and the retarded-time construction}

\subsection{The Kerr theorem}

Traditional formulation of the Kerr theorem is following.

Any  geodesic and shear-free null congruence in
Minkowski space is defined by a function $Y(x)$ which is a solution of the
equation
\be         F  = 0 ,                            \label{(1.1)}\ee
where   $F (\l_1,\l_2,Y)$   is an arbitrary analytic
function of the projective twistor coordinates
\be
Y,\qquad \l_1 = \z - Y v, \qquad \l_2 =u + Y \Z .\label{(1.2)}
\ee
The vector field
\be
 e^3 = du+ \Y d \z  + Y d \Z - Y \Y d v  = P k_\m dx^\m
\label{1.8}
\ee
determines the congruence then
in the null cartesian coordinates
$u, \ v, \ \Z , \ \z $.\fn{The field $k_\m$ is a normalized form of
$e^3_\m$ with $k_\m \Re e \dot x_0^\m=1$.}

In the Kerr-Schild backgrounds
the Kerr theorem acquires a more
wide contents \cite{DKS,IvBur1,BKP}.
It allows one to obtain the position of singular lines, caustics of the
PNC, as a solution of the system of equations
\be F=0;\quad \d F /\d_Y =0 \ , \label{sing}\ee
and to determine
the important parameters of the corresponding solutions:
\be
\tilde r = - \quad d F / d Y , \label{tr}
\ee
and
\be
P = \d_{\l_1} F - \Y \d_{\l_2} F.  \label{PF}
\ee
Parameter $\tilde r$ characterizes a complex radial distance,
and for the stationary Kerr solution it is a typical complex
combination $\tilde r= r+ia \cos\theta$. Parameter $P$ is connected with
the boost of source.
For details we refer reader to \cite{Bur-nst}.

Working in $CM^4$ one has to consider $Y$ and $\bar Y$ functionally
independent, as well as the null coordinates $\z $ and  $\Z$. Coordinates
$u,v$ and congruence turn out to be complex.  The
 corresponding complex null tetrad (\ref{Kt}) may be considered as a basis of
$CM^4$.  The Kerr theorem determines in this case only the ``left" complex
structure - function $F(Y)$. The real congruence appears as an
intersection with a complex conjugate ``right'' structure.

\subsection{Quadratic function F(Y) and interpretation of parameters.}
It is instructive to consider first stationary case.
Stationary congruences having Kerr-like singularities contained in
a bounded region have been considered in papers \cite{IvBur,Bur1,KerWil}.
It was shown that in this case function $F$ must be at most quadratic in $Y$,
\be
F \equiv a_0 +a_1 Y + a_2 Y^2 + (q Y + c) \l_1 - (p Y + \q) \l_2,
\label{FK}
\ee
where coefficients $ c$ and $ p$ are real constants
and $a_0, a_1, a_2,  q, \q, $  are complex constants.
Killing vector of the solution is
determined as
\be
\hat K = c\d _u + \q \d _\z + q \d _\Z -p\d _v . \label{Knull}
\ee
Writing the function F in the form
\be
F = A
  Y^2 + B Y + C, \label{Fquadr}
\ee
one can find two solutions of the equation $F=0$ for the
function $Y(x)$
\be
 Y_{1,2} = (- B \pm \D )/2A, \label{Y12}
\ee
where $ \D = (B^2 - 4AC)^{1/2}.$

On the other hand from  (\ref{tr})
\be
\tilde r = - \d F /\d Y= -2AY -B,
\label{tr2}
\ee
and consequently
\be
\tilde r =PZ^{-1} = \mp \D.
\label{tr1}
\ee
Two roots reflect the known twofoldedness of the Kerr
geometry.  They correspond to two different directions of
congruences on positive and negative sheets of the Kerr space-time.
The expression (\ref{PF}) yields
\be
P=pY\Y  + \q \Y + qY +c \ . \label{Ppc}
\ee
\subsection{Link to the complex world line of source.}
The stationary and boosted Kerr geometries are described
by
a  straight complex world line with a real 3-velocity $\vec v$ in $CM^4$:
\begin{equation}
x_0^\m (\t) = x_0^\m (0) + \xi^\m \t; \qquad \xi^\m = (1,\vec v)\ .
\label{dec}
\end{equation}
The gauge of the complex parameter $\t $ is chosen
in such a way that $Re \ \tau$ corresponds to the real time $t$.

The quadratic in $Y$ function $F$ can be expressed in this case in the form
\cite{IvBur,Bur1,BKP,BurMag}
\begin{equation}
F \equiv (\l_1 - \l_1^0) \hat K\l_2 - (\l_2 -\l_2^0) \hat K\l_1 \ ,
\label{Fx0}
\end{equation}
where the twistor components $\l_1, \ \l_2 $ with zero indices
denote their values
on the points of the complex world-line $ x_0 (\t)$, (\ref{np}),
and $\hat K$ is a Killing vector of the
solution
\begin{equation}
\hat K = \d _\tau x_0^\m(\t) \d_\m = \xi^\m \d_\m \ . \label{hK}
\end{equation}
Application $\hat K$ to $\l_1$ and $\l_2$ yields the expressions
\bea
\hat K \l_1 &=& \d _\tau x_0^\m(\t) e^1_\mu ,
\nonumber\\
\hat K \l _2 &=&\d _\tau x_0^\m (e^3_\m-\Y e^1_\m ).
\label{cKll}
\eea
From  (\ref{PF}) one obtains in this case
\be
P= \hat K \rho = \d _\tau x_0 ^\mu (\tau)e^3_\mu \ ,
\label{Prho}
\ee
where
\be
\rho= \l_2 + \Y \l_1 = x^\mu e^3_\mu \label{rho},
\ee
Comparing  (\ref{Prho}) and
  (\ref{Ppc}) one obtains the correspondence in
 terms of $ \ p,\ c,\ q, \ \q $,
\be
\hat K\l _1= pY+\q, \qquad \hat K \l _2=qY +c,
\label{Kll}
\ee
that allows one to set the relation between parameters
$p, c, q, \q$ and  $\xi ^\mu$
showing that these parameters are connected with the boost of the source.

The complex initial position of complex world line $x_0^\m(0)$  in
(\ref{dec}) gives six more parameters to solution, which are connected with
coefficients $a_0, \ a_1 \ a_2 \ $.
It can be decomposed as $\vec x_0 (0) = \vec c + i\vec d$, where $\vec c $ and
$\vec d$ are real 3-vectors with respect to the space O(3)-rotation.
 The real part, $\vec c$, defines the initial position of
source, and the imaginary part, $\vec d$, defines the value and direction of
angular momentum (or the size and orientation of singular ring).

It can be easily shown that in the rest frame, when $\vec v=0, \quad \vec d
=\vec d_0 $, the singular ring lies in the plane orthogonal to $\vec d$
and has a radius $a=\vert \vec d_0 \vert $.
 The corresponding angular momentum
is $\vec J = m \vec d_0.$

\subsection{L-projection and complex retarded-time parameter.}

In the form (\ref{FK}) all the coefficients are constant while the
form (\ref{Fx0}) has an extra explicit linear dependence on $\tau$
via terms $ \l_1^0 (x_0(\t))$ and $ \l_2^0 (x_0(0))$.
However,
this dependence is really absent. As consequence of
the relations $ \l_1^0 (x_0(\t)) = \l_1^0 (x_0(0)) + \t \hat K \l_1,\quad
\l_2^0 (x_0(\t)) = \l_2^0 (x_0(0)) + \t \hat K \l_2 $ the terms proportional
to $ \t$ cancels and these forms are equivalent.

Parameter $\t$ may be defined for each point
$ x$ of the Kerr space-time and plays the role of a complex retarded time
parameter.  Its value for a given point $x$ may be defined by
L-projection, using the solution $Y(x) $ and forming the twistor parameters
$ \l_1,\quad \l_2 $ which fix a left null plane.

$L$-projection of the point $x$ on the complex world line $x_0(\t)$ is
determined by the condition
\be (\l_1-\l_1^0)|_L = 0,\qquad (\l_2-\l_2^0)|_L
=0 \ , \label{Lnp} \ee
where the
sign $|_L$ means that the points $x$ and $x_0(\t)$ are synchronized by the
left  null plane (\ref{L}),
\be x- x_0(\t_L) = \alpha e^1 + \beta e^3 . \ee
The condition (\ref{Lnp}) in representation
(\ref{lambx}) has the form
\be
 (x^\m -x_0^\m) e^1_\m |_L =0, \qquad  (x^\m -x_0^\m)
(e^3_\m - \Y e^1_\m)|_L=0,
\label{ll0}
\ee
which shows that the points $x^\m $
and $x_0^\m$ are connected by the left null plane spanned by null
vectors $e^1$ and $e^3$.

This left null plane belongs simultaneously to the "in" fold of the
light cone connected to the point $x$  and to the "out" fold of the light
cone emanating from point of complex world line $x_0$.  The point of
intersection
of this plane with the complex world-line $x_0(\t)$ gives a value of the
"left" retarded time $\t_L$, which is in fact a complex scalar function on
the (complex) space-time $\t_L(x)$.

By using the null plane equation (\ref{Lnp}) one can  express $\D$ of
(\ref{tr1}) in the form
\be \Delta |_L= (u-u_0) \dot v_0 + (\z - \z_0 ) \dot
{\Z _0} + (\Z - \Z_0) \dot \z_0 + (v - v_0) \dot u_0 =\frac 12 \d_\t (x - x_0
)^2 = \t _L -t + \vec v \vec R , \label{Dpm}\ee
where
\be \vec v = \dot {\vec
x_0}, \qquad \vec R = \vec x - \vec x_0. \label{3.23}\ee

It gives a retarded-advanced time equation
\be \t = t \mp \tilde r + \vec v  \vec R,
\label{ret-adv}\ee
and a simple expression for the solutions $Y(x)$
\be Y_1 =  [ (u-u_0) \dot v_0 + (\z - \z_0) \dot{\Z}_0]
/ [ (v - v_0) \dot {\Z}_0 - (\Z -\Z_0) \dot v_0],
\ee
and
\be Y_2 =
[ (u-u_0) \dot \z _0 - (\z - \z_0) \dot{u}_0]/
[ (u-u_0) \dot v_0 + (\z - \z_0) \dot{\Z}_0] .
\ee

For the stationary Kerr solution $\tilde r=r+ia\cos\theta$,
and one sees that the second root $Y_2(x)$ corresponds to a transfer to
negative sheet of metric: $r\to -r; \quad \vec R \to -\vec R$ with a
simultaneous complex conjugation  $ia \to -ia$.

Introducing the corresponding operations:
\be
P: r\to -r, \quad \vec R\to - \vec R \label{P}
\ee
\be
C: x_0 \to \bar x_0, \label{C}
\ee
and  also the transfer $"out" \to "in" $
\be
T: t-\tau \to \tau -t. \label{T}
\ee
One can see that the roots and corresponding Kerr congruences are
CPT-invariant.

\subsection{Nonstationary case.  Real slice.}

In nonstationary case this construction acquires new peculiarities:

i/ coefficients of function $F$ turn out to be variable and dependent on
a retarded-time parameter,

ii/ $\d _\tau x_0^\mu =\xi ^\mu $ can take complex values, that implies
complex values for function $P$ and was an obstacle for obtaining the real
solutions \cite{BKP}.

iii/ $K$ is not Killing vector more.

 To form the real slice of space-time, we have to consider, along with
the ``left'' complex structure, generated by a ``left'' complex world line
$x_0$, parameter $Y$ and by the left null planes, an independent ``right''
structure with  the ``right'' complex world line $\bar x_0$, parameter $\Y$
and the right null planes, spanned by $e^2$ and $e^3$.
These structures can be
considered as functionally independent in $CM^4$, but they have to be
 complex conjugate on the real slice of  space-time.

First note, that for a real point of space-time $x$ and for the corresponding
real null direction $e^3$,  the values of function
\be \rho (x) = x^\m e^3_\m(x)
\label{rhonst}
\ee
are real.
Next, one can determine the
values of $\rho$ at the points of the left and right complex world lines
$x_0^\mu$ and $\bar x_0 ^\mu$ by L- and R-projections
\be \rho _L (x_0) =
x_0^\m e^3_\m (x)|_L,\label{rhoL}\ee
and
\be \rho _R(\bar x_0) = \bar x_0^\m
e^3_\m (x)|_R,\label{rhoR}\ee
For the "right" complex structure, points $x$ and $\bar x_0(\bar \t)$
are to be synchronized by the right null plane
$ x - \bar x_0(\bar \t _R) = \alpha e^2 + \beta e^3 $.
As a consequence of the
conditions $e^{1\mu} e^3_\mu=e^{3\mu} e^3_\mu =0$, we obtain
\be \rho _L(x_0) = x_0^\m e^3_\m (x)|_L = \rho (x).
\label{rho0}\ee
So far as the parameter
$\rho (x) $ is real, parameter $\rho _L (x_0)$ will be real, too.
Similarly,
\be \rho _R(\bar x_0) = {\bar x_0}^\m e^3_\m (x)|_R = \rho (x),
\ee and consequently, \be \rho _L(x_0) = \rho (x) = \rho _R (\bar
x_0).
\label{(4.14)}\ee
By using  (\ref{lambx}) and (\ref{rhonst}) one obtains
\be \rho = \l_2 + \Y
\l_1.  \label{rhotw} \ee
Since L-projection (\ref{Lnp}) determines the  values of the
left retarded-time parameter $\tau _L= (t_0 + i\sigma)|_L$,
the real function $\rho$ acquires a dependence on the
retarded-time parameter $\tau _L$.
 It should be noted that
the real and imaginary parts of $\tau |_L$
are not independent because of the constraint caused by
L-projection.

It means that the real functions $\rho $ and
 $\rho _0 $ turns out
to be functions of real retarded-time parameter
 $t_0=\Re e \  \tau _L$, while
 $\l^0_1$ and $\l^0_2$ can also depend on
$\sigma $.

These parameters are constant on the
left null planes that yields the relations
\be
(\sigma |_L),_2 =(\sigma |_L),_4 = 0 , \quad
(t_0|_L),_2 =(t_0|_L),_4 = 0 \ . \label{st02}
\ee
In analogue with the above considered stationary case,
one can restrict function $F$ by the quadratic in $Y$ expression
\be
F \equiv (\l_1 - \l_1^0) K_2 - (\l_2 -\l_2^0) K_1  ,
\label{Fnst}
\ee
It can be shown \cite{Bur-nst} that
the functions $K_1$ and $K_2$ are linear in $Y$ and depend on the
retarded-time $t_0$. It
leads to the form (\ref{FK}) which coefficients shall depend on the
retarded-time
\be
 K_1 (t_0) =\d_{t_0} \l_1^0 ,\qquad K_2 (t_0) =\d_{t_0} \l_2^0 \ .
\label{K12} \ee
In tetrad representation (\ref{lambx}) it takes the form
\be
 K_1 =\d _{t_0} x_0^\m e^1 _\m ,\qquad
K_2 =\d _{t_0} x_0^\m (e^3 _\m -\Y e^1 _\m),
\label{K12t} \ee
and
\be
P=\Y K_1 +K_2 ,
\ee
that yields for function $P$ the real expression
\be
P=\d _{t_0} (x_0^\m e^3_\m)|_L = \d _{t_0} \rho _L \ .
\label{Pnst}
\ee
It is seen that $\rho(t_0)=\rho _L(t_0)$ plays
the role of a potential for $P$, similarly to some nonstationary solutions
presented in \cite{KraSte}.

It seems that the extra dependence of function $F$ from  the non-analytic
retarded-time parameters $t_0$  contradicts to the Kerr theorem,
however the non-analytic part disappears by L-projection and analytic
dependence on $Y,\quad l_1, \quad l_2$ is reconstructed.
Note, that all the {\it real} retarded-time
derivatives on the real space-time are non-analytic and have to involve the
conjugate right complex structure. In particular, the expressions
(\ref{K12t}) acquire the form
\be
 K_1 =e^1 _\m \Re e \dot x_0^\m  ,\qquad
K_2 =(e^3 _\m -\Y e^1 _\m)\Re e \dot x_0^\m  ,
\label{K12dot} \ee
where $\dot x_0^\m = \d _{t_0} x_0^\m \ .$

\section{Field equations}
\bigskip
Field equations for Einstein-Maxwell system in the Kerr-Schild class
were obtained in \cite{DKS}.
Electromagnetic field is given by tetrad components of selfdual tensor
\be \cF _{12} =AZ^2 \label{1}\ee
\be \cF _{31} =\gamma Z - (AZ),_1  \ . \label{2}\ee
The equations for electromagnetic field are
\be A,_2 - 2 Z^{-1} \cZ Y,_3 A  = 0 , \label{3}\ee
\be \cD A+  \cZ ^{-1} \gamma ,_2 - Z^{-1} Y,_3 \gamma =0 .
\label{4}\ee
Gravitational field equations are
\be M,_2 - 3 Z^{-1} \cZ Y,_3 M  = A\bar\gamma \cZ ,  \label{5}\ee
\be \cD M  = \frac 12 \gamma\bar\gamma  , \label{6}\ee
where
\be
\cD=\d _3 - Z^{-1} Y,_3 \d_1 - \cZ ^{-1} \Y ,_3 \d_2   \ . \label{cD}
\ee
Solutions of this system were given in \cite{DKS} only for stationary case
for $\gamma=0$.
In this paper we give a preliminary analysis of the nonstationary
solutions for $\gamma \ne 0$.
The principal new point is the existence of retarded-time
parameter $t_0$, which satisfies
\be
(t_0),_2 =(t_0),_4 = 0 \ . \label{7}
\ee
The equation (\ref{3}) takes the form
\be
(AP^2),_2=0 \ , \label{8}
\ee
and has the general solution
\be A= \psi(Y,t_0)/P^2.
\label{9}\ee

Action of operator $\cD$ on the variables $Y, \bar Y $ and $ \rho$
is following \be \cD Y = \cD \bar Y = 0,\qquad \cD \rho =1 \ .
\label{10}\ee From these relations and (\ref{Pnst}) we have $\cD
\rho = \d \rho / \d t_0 \cD t_0  = P\cD t_0 =1 $, that yields \be
\cD t_0 = P^{-1} . \ee As a result the equation (\ref{4}) takes
the form \be \dot A = -(\gamma P),_{\bar Y} , \label{11} \ee where
$\dot {( \ )} \equiv \d_{t_0}$. Note, that the derivatives $\d _Y$
and $\cD$ commute \be \cD \d _Y - \d _Y \cD =0 , \ee that allows
to obtain by integration \be \gamma P =  - \d_{t_0} (\psi \ln P +
\phi (Y) )/ P_{\bar Y}. \label{12}\ee The equations (\ref{5})and
(\ref{6}) take the simple form \be m,_{\bar Y}=  P^3 A\bar\gamma ,
\label{13}\ee and \be \frac 1P \dot M= \frac 12 \gamma\bar\gamma .
\label{14}\ee We consider now the simplest solution corresponding
to the case  $\dot A =\dot P=0$  without oscillations of the ring.
In this case $\gamma = \chi (Y, t_0) /P$ that describes a null
electromagnetic radiation  $\cF _{31}= \gamma Z -(AZ),_1 $
propagating along the Kerr PNC direction $e^3$. In accordance with
(\ref{14}) it has to lead to a loss of mass by radiation with the
stress-energy tensor $\kappa T_\mn = \frac 12 \gamma \bar \gamma
e^3_{\m} e^3_{\n}$, similarly to the Vaidia ``shining star''
solution \cite{KraSte,VaiPat}. However, the Kerr twofoldedness
shows us that the loss of mass on the positive sheet of metric is
really compensated by an opposite process on the negative sheet
with an in-flow of the radiation. The structure of the Kerr PNC
shows that there are no discharges or sources of this flow. One
can assume that the $\gamma$ is really a vacuum zero-point field,
or the field of the null vacuum fluctuations, which has a
resonance on the Kerr singular ring. The zero-point field has a
semi classical nature since it has a quantum origin, but a
classical exhibition in the form of the well known Casimir effect.
Therefore, this radiation can be interpreted as a specific
exhibition of the Casimir effect.  To improve the situation with
the loss of mass by this field one can use the known recipe for
computation of the Casimir effect \cite{deWit}. The classical
energy-momentum tensor has to be regularized \be T_{reg}^\mn = \
:T^\mn: \ \equiv T^\mn -<0|T^\mn|0> \ee under the condition
$\nabla _\m T_{reg}^\mn =0$. On the classical level of the
 Einstein-Maxwell equations, this procedure corresponds exactly to the
subtraction of the term $\gamma\bar\gamma$ from (\ref{14}).

Since this radiative term $\gamma\bar\gamma$ will also appear in all other
oscillating solutions,
it is tempting to conjecture that on the quantum level this procedure is
equivalent to the postulate on the absence of radiation by oscillations.

Preliminary treatment of the other oscillating solutions shows one more
peculiarity of this system: the appearance of an imaginary contribution to
the mass parameter. Such contributions are not admissible in the standard
Kerr-Schild formalism, but the imaginary mass term appears in some other
formulations, in particular,  in the form of the NUT-parameter of the
Kerr-NUT solution.
By a nonzero value of this parameter there appears one more
`axial' singular filament (Dirac monopole string) which threads the
Kerr singular ring and is extended to infinity.
Note, that such an `axial' string was also observed in the supersymmetric
extension of the Kerr-Newman solution by the treating the fermionic traveling
waves \cite{BurSup}. As it was mentioned in \cite{BurSup},
this second singular filament, being topologically coupled to
the Kerr singular ring, acquires an interesting
 interpretation as a carrier of the de Broglie wave. In
the space-time with nontrivial boundaries this filament can
play the role of a guide in a topological wave-pilot construction.

More detailed analysis of the system of these equations will be given
elsewhere.

\section*{Acknowledgments}
Author thanks Organizing Committee for very kind invitation
and financial support and also V. Kassandrov and B.Frolov for useful
discussions.

\section*{Appendix A. Basic relations of the Kerr-Schild formalism}
Following the notations of the work
\cite{DKS},  the Kerr-Schild null tetrad
 $e^a =e^a_\m dx^\m $ is determined by relations:
\begin{eqnarray}
e^1 &=& d \zeta - Y dv, \qquad  e^2 = d \bar\zeta -  \bar Y dv, \nonumber \\
e^3 &=&du + \bar Y d \zeta + Y d \bar\zeta - Y \bar Y dv, \nonumber\\
e^4 &=&dv + h e^3,\label{Kt}
\end{eqnarray}
and
\be g_{ab}= e_a^\m e_{b\m} = \left(
\begin{array}{cccc} 0&1&0&0 \\ 1&0&0&0 \\ 0&0&0&1 \\ 0&0&1&0 \end{array}
\right). \label{gab} \ee
Vectors
$e^3, e^4$ are real, and $ e^1, e^2 $ are
complex conjugate.

 The Ricci rotation coefficients are
given by \be \G ^a_{bc} = - \quad e^a_{\m;\n} e_b^\m e_c^\n.  \label{(1.4)}
\ee
 The PNC have the $e^3$  direction as tangent.  It will be geodesic if and
only if $\G_{424} = 0$ and shear free if and only if $\G_{422} = 0$.  The
 corresponding complex conjugate terms are $\G_{414} = 0$ and $\G_{411} = 0$.

The inverse (dual) tetrad has the form
 \bea
  \d_1 &=& \d_\z  - \Y \d_u ;
\nonumber\\
\d_2 &=&  \d_\Z - Y \d_u ;
\nonumber\\
 \d_3 &=&  \d_u - h \d_4  ;
\nonumber\\
 \d_4 &=&  \d_v + Y \d_\z + \Y \d_\Z - Y  \Y \d_u ,  \label{1.10}
\eea
where $\d _a \equiv ,_a \equiv e_a^\m \d ,_\mu $.

Parameter $Z=Y,_1 =\rho +i \omega$ is a complex expansion of congruence
$\rho=expansion$ and $\omega = rotation$. $Z$ is connected with a complex
radial distance $\tilde r$ by relation
\be
\tilde r =PZ^{-1}. \label{trZ}
\ee

It was obtained in \cite{DKS} that connection forms in Kerr-Schild metrics
 are
\be \G_{42} = \G_{42a} e^a  = - d Y - h Y,_4
e^4 .  \label{1.11}
\ee
The congruence  $e^3 $ is geodesic if $ \G_{424} =
-Y,_4 (1-h) = 0, $ and is shear free if $ \G_{422} = -Y,_2 = 0.$
Thus,  the function $ Y (x)$ with conditions
\be
Y,_2 = Y,_4 = 0,  \label{1.12}
\ee
defines a shear-free and geodesic congruence.

\end{document}